\documentclass{article}
\usepackage[utf8]{inputenc}
\pdfoutput=1
\usepackage{amsmath,amsthm,amsfonts}
\usepackage{mathtools}
\usepackage{xcolor}
\usepackage{dirtytalk}

\newtheorem{proposition}{Proposition}
\newtheorem{observation}{Observation}
\newtheorem{conjecture}{Conjecture}

\theoremstyle{definition}
\newtheorem{example}{Example}
\newtheorem{remark}{Remark}
\newtheorem{definition}{Definition}

\title{Electing the Executive Branch}

\author{
Rutvik Page \\
London School of Economics and Political Science \\
r.r.page@lse.ac.uk
\and
Ehud Shapiro \\
Weizmann Institute \\
ehud.shapiro@weizmann.ac.il
\and
Nimrod Talmon \\
Ben-Gurion University \\
talmonn@bgu.ac.il
}

\date{}

\begin{document}

\maketitle

\begin{abstract}

The executive branch, or government, is typically not elected directly by the people, but rather formed by another elected body or person such as the parliament or the president.  As a result, its members are not directly accountable to the people, individually or as a group.  We consider a scenario in which the members of the government are elected directly by the people, and wish to achieve proportional representation while doing so.
We propose a formal model for allocation of $k$ offices (ministries), each associated with a disjoint set of candidates contesting for that seat; a set of voters provide approval ballots for all offices. We identify various scenarios pertaining to the `power' exercised by different offices and propose axioms and algorithms to satisfy these axioms to determine  proportionally representative office allocations. 
As using a simple majority vote for each office independently might result in disregarding minority preferences altogether, here we consider an adaptation of the greedy variant of Proportional Approval Voting (GreedyPAV) to our setting, and demonstrate---through computer-based simulations---how voting for all offices together using this rule overcomes this weakness and satisfies a proportionality axiom.
We note that the approach is applicable also to a party that employs direct democracy, where party members elect the party's representatives in a coalition government. To this effect, we propose a simply explainable pen-and-paper realisation of GreedyPAV that is resilient in situations where candidates elected to a certain office choose not to fill it. 
\end{abstract}

\section{Introduction}

Consider a scenario in which a government in a country has to be populated; i.e., there should be elected members of the government like the minister for health, the minister for education, etc. Usually this assignment process is done via a non-participatory process.
In this paper, we describe a way of doing this process in a participatory way. In essence, we would like each of the citizens of the country to describe their preferences regarding the assignment of alternatives to each office.

While the above setting is indeed quite imaginary, as this is not a generalised practice, our particular motivation for this work comes from populating governments in coalitional systems; indeed, in coalitional systems, following coalition negotiations, each party in the coalition is being allocated some set of offices which, in turn, have to be populated with ministers.
Specifically, we are interested in the process in which a party that got allocated some offices through such a negotiation, shall  decide internally---via democratic vote of its members--a way to assign ministers to each of their allocated offices.

We view this process as a social choice setting and first observe that one natural and simple way to approach it is to view it as $k$ independent elections, where $k$ is the number of offices allocated to the party. For example, each voter can select a set of alternatives for each of the offices and, for each office independently, we can select the alternative that got the highest number of votes.

Observe, however, that such a process might disregard the preferences of minorities altogether; in particular, if there an election by the way of strict majority votes for some alternatives for each of the offices independently, then only the alternatives voted by the majority would be selected at every office, and none of the alternatives of the minority would be, even if the minority consist of $50 - \epsilon$ percent of the votes.

To overcome this weakness of total minority exclusion, we view the setting as a whole, in which we have one election whose output would be the complete assignment of alternatives to the full set of $k$ offices, and this assignment shall be approximately, if not fully proportional.

We start by defining two classes of axioms for a special case of our setting and generalise it voter expressivity-wise; in essence, we allow for voters to express their broader opinions about the relative `power' of the offices over each other. In each of these settings, we propose a range of proportionality axioms and propose aggregation methods to realise committees that adhere to these axioms. During this course, we remark that our special settings are a domain restriction of our more general settings. Towards to end, we study a specific rule, GreedyPAV, in our settings and show -- via computer-based simulations, with two different ways of synthetic preference generation -- that it guarantees proportional representation to minorities in many cases. Further, we also propose a Paper-and-Pencil Realization of the GreedyPAV rule that makes it much more accessible in practical settings.

\section{Related Work}




To the best of our knowledge, the specific setting we are considering in this paper hasn't been discussed before.
The most related work is that of Conitzer et al.~\cite{conitzer2017fair}, who study a generalization of our model and study proportionality and a varied range of fairness axioms assuming cardinal preferences. They consider the case of public decision making on a set of issues, each associated with a disjoint set of possible solutions and every voter has a utility function which is solution specific. Their settings assume the number of issues to vastly outnumber the number of voters unlike in our settings, where the number of offices to be filled can only be limited and the number of voters can be large.

There are some related models studied in the social choice literature, however, that we mention below.

First, 
we mention the work of Boehmer et al.~\cite{boehmer2020line} that considers an assignment social choice problem, but differs from our model in that voters provide numerical utilities and alternatives can run for few offices in parallel (so the output decision shall take into account the suitability of alternatives to offices, while we derive the suitability directly for the votes).

Generally speaking, our social choice task is of selecting a committee, and thus is related to the extensive work on committee selection and multiwinner elections~\cite{mwchapter}. In our setting, however, we do not aim simply at selecting $k$ alternatives, but at selecting an assignment to $k$ offices.

Related, the line of work dealing with committee selection with diversity constraints (e.g., see~\cite{izsak2017working,aziz2019rule}) has some relation to our work, in particular, as one can choose a quota of ``at most one health minister'', ``at most one education minister'', and so on.

Lackner~\cite{Lackner_2020} propose the concept of \emph{perpetual voting} which lays a framework that helps circumvent the difficulties faced by single shot decisions making by enabling minorities to have a \say{fair (proportional)} say over the decision making process. This in turn ensures an inclusive decision making process by incentivising minority voters to participate with the hope of not being overlooked in the long term.

Talmon et al.~\cite{repeated} study a similar temporal setting, though their point of focus differs from that of perpetual voting. While the latter proposes several voting rules and examines them from the purview of three axiomatic properties (which in turn cater to the representation of individual voters), the former concentrates on the generalization of popularly known proportional representation~\cite{aziz2017justified,sanchez2017proportional} axioms to different scenarios in terms of preference elicitation and extent of cohesiveness amongst deserving voter groups.

\section{Formal Model}

  Suppose, we have a set of $k$ offices and $k$ corresponding disjoint sets of alternatives, $A_j$, $j \in [k]$ so that each candidate runs for at most one office). Let us denote the set of all alternatives by $A := \cup_{i \in [k]} A_j$.
  Here we consider the approval ballots, thus, we have a set $V = \{v_1, \ldots, v_n\}$ of $n$ votes such that $v \subseteq A$, $\forall v \in V$.
An aggregation method for our setting takes as input such an instance $(A, V)$ and outputs a set, \emph{Office Allocation}, $X = \{w_1, \dots, w_k\}$, where $w_i \in A_i$, $\forall i \in [k]$, which signifies the winners for the corresponding offices. Define a function $eligible: O \rightarrow 2^A \setminus \emptyset$ such that $eligible(o)$ denotes the set of candidates that are eligible for contesting election for some office $o \in O$. A \emph{Power Function}, $\mathbf{P} : V \times O \longrightarrow \mathbb{R}$ denotes the amount of influence that a voter $v \in V$ believes that an office $o \in O$ has. In essence, $P(i, o)$ (or $P_i(o)$) is the amount of power that a voter $v \in V$ believes that an office $o \in O$ wields. We assume that power functions for all voters and corresponding offices are non-negative and additive; that is, $P(i, o) \geq 0$, $\forall i \in V$, $o \in O$ and the power realised by a voter $i$ for an office allocation $X$ is $\Sigma_{o} P(i, o)$, $\forall o \in A_i \cap X$.

\begin{example}\label{example:toy}
Consider the $3$ sets of alternatives to be $A_1 = \{a, b\}$, $A_2 = \{c, d\}$, and $A_3 = \{e, f, g\}$, and the set of votes $v_1 = \{a, c, e\}$, $v_2 = \{a, c, f\}$, and $v_3 = \{a, d, f\}$.
An output of an aggregation method might be an office allocation, $X = \{a, c, g\}$, corresponding to alternative $a$ being selected for the first office, alternative $c$ being selected for the second office, and alternative $g$ being selected for the third office. Note that since a candidate contests election only for a single office, an office allocation can be a set rather than a tuple. 
\end{example}

\section{The Problem with Independent Elections}

Perhaps the most natural and simple solution would be to view the setting as running $k$ independent elections; for example, selecting to each office the alternative that got the highest number of approvals.
This, however, would be problematic; in particular it would not be proportional.

\begin{example}\label{example:one}
Consider a society with strict majority voting for $a_j \in A_j$ for each $j \in [k]$. Now, disregarding how the other voters vote, $a_j \in A_j$, $j \in [k]$ would be selected.
In particular, even a minority of $49\%$ would not be represented in the government.
\end{example}

To aid this disregard of minorities in the elections, we propose fairness axioms (and algorithms to achieve office allocations that follow these fairness axioms) that seek to promote a proportional representation to voter who vote for electing the executive.

\section{Fairness Axioms}
We formally propose fairness axioms in this section and corresponding algorithms that find office allocations that satisfy these fairness axioms. 

\subsection{Global Axioms}

\begin{definition}[(Global Justified Representation)]
An office allocation $X$ is said to satisfy \emph{Global Justified Representation} if $\forall V' \subseteq V$ such that $|V'| \geq n / k$, and  $\forall j \in [k]$,
$$((\cap_{v \in V'} v) \cap A_j) \neq \emptyset \implies (X \cap (\cup_{v \in V'} v)) \neq \emptyset$$

\end{definition}\begin{definition}[Global Proportional Justified Representation]
An office allocation $X$ is said to satisfy \emph{Global Proportional Justified Representation} if $\forall V' \subseteq V$ and $\forall j \in [k]$
$$((\cap_{v \in V'} v) \cap A_j) \neq \emptyset \implies  |(X \cap (\cup_{v \in V'}v))| \geq \lfloor \frac{|V'|}{|V|} k \rfloor $$
\end{definition}

Global Justified Representation mandates that at the end of any voting rule every voter group of size $\geq n / k$ who have a non empty intersection of approval sets for all offices should get one of their preferred candidates on at least one of all available offices.

\begin{proposition}\label{proposition:GJR}
A committee satisfying \emph{Global Justified Representation} always exists and can be calculated in polynomial time. 
\end{proposition}
\begin{proof}
The algorithm to calculate a committee satisfying \emph{Global Justified Representation} progresses in $k$ steps as follows: if there is any candidate who is approved by all the voters, she can be made to assume any of the offices and the axiom is trivially satisfied. If not so, in every iteration, we find the approval score of each and every candidate and select the one with the highest approval score $\geq n/k$ into one of the offices, subsequently exclude the satisfied voters from further consideration and reiterate. At any point in the algorithm, if none of the candidates have an approval score $\geq n / k$, we ask random unelected candidates to assume the empty offices. \\
For correctness, observe that in each iteration of the algorithm, disjoint sets of $\geq n / k$ voters are satisfied. So, as long as there are disjoint voter groups of size $ \geq n / k$, the algorithm satisfies those voter groups by providing office to one of their favorite candidates. Since there can be only $\leq k$ disjoint voter sets each of size $n / k$, our axiom satisfies $k*(n / k)$ i.e. $n$ voters in at most $k$ iterations.  
\end{proof}

\begin{conjecture}
GreedyPAV satisfies GJR.
\end{conjecture}
\begin{proof}
We base this conjecture on the basis of our simulations, which are presented later in the paper. 
\end{proof}

Proposition \ref{proposition:GJR} establishes that an office allocation that satisfies GJR always exists and can be computed in polynomial time. While GJR takes care of both deserving as well as cohesive groups, it does not provide sufficient representation when for example, the voter group size is $5 \cdot n / k$ and it still provides office to only one of their preferred candidates instead of at least $5$. In such cases, the number of candidates that are added to offices for satisfying larger voter sets should be proportional to the size of the voter group. GPJR mitigates this short coming of GJR by providing proportional representation to voter groups for whom representation is justified (i.e. deserving and cohesive voter groups).

We provide a super-polynomial time algorithm that satisfies Global Proportional Justified Representation. 

\begin{proposition}{\label{proposition:GPJR_algo}}
An office allocation satisfying Global Proportional Justified Representation always exists.
\end{proposition}

\begin{proof}
 Consider the power-set of the voter set (having $n$ voters), the size of which is $2^n$. Then, for each $V' \in \mathcal{P}(V)$, we check if the intersection of the approval ballots of every $v \in V' \neq \emptyset$ on every office $o \in O$, and if so, we add $v$ to $V''$. Thus, $V''$ is that set of subset of voters who agree on at least one candidate on all the offices. Formally, $$V'' = \{V' \subseteq V: \forall i,j,k, v_i \cap v_j \cap A_k \neq \emptyset \}$$
We cycle through $V''$, picking up the largest group of voters $V'$ in each iteration and allow $\lfloor \frac{|V'|}{|V|} \rfloor$ of the voter group's favourite candidates to assume offices. Further, eliminate all those groups of voters from $V''$ which have a non empty intersection with $V'$, i.e. we eliminate all those voter groups those who have been catered to by satisfaction of one of their fellow voters. We terminate when there is either no voter group left to satisfy or when the size of the largest voter group is $< n/k$. 
For correctness, notice that once the algorithm takes care of a voter subset $V'$ (of size $\geq n/k$) then, $\forall v \in V'$, it eliminates every single voter group that has $v$ in it. Thus, only disjoint sets of voters of size $\geq n/k$ are \emph{proportionally} satisfied. Due to exactly this, the algorithm does not run out of number of offices to be filled. Moreover, we satisfy groups of voters in a decreasing order of their size because this allows the pruning of other voter groups that have a non-empty intersection with them while providing them with at least proportional representation. \\
Hence, an office allocation satisfying GPJR always exists, albeit using a super polynomial time algorithm. 
\end{proof}
\begin{proposition}
Global Proportional Justified Representation is fixed-parameter tractable wrt. the number of voters $n$. 
\end{proposition}
\begin{proof}
Note that the algorithm mentioned in Proposition $\ref{proposition:GPJR_algo}$ cycles through the super-set of the set of voters whose size is $2^n$, hence the parameterization by the number of voters $n$. For each voter subset $V'$, we check if every pair of voters have a non-empty intersection on every office $k$ in time $\mathcal{O}(n^2\cdot k)$ and add it to the set $V''$ if the condition is satisfied. In turn, the algorithm cycles through the set $V''$ (the size of which is also a function of $n$) and eliminates voter subsets from it till one of the termination conditions are not met. This procedure in effect has a running time depending only on the number of voters $n$. Therefore, the algorithm selecting an office allocation that satisfies Global Proportional Justified Representation is indeed fixed-parameter tractable with respect to the number of voters $n$. 
\end{proof}
Therefore, an office allocation satisfying Global Proportional Justified Representation (and hence, Global Justified Representation) always exists. However, while GJR can be satisfied using a polynomial time procedure, there exists a super-polynomial time algorithm that satisfies GPJR and is FPT with respect to the number of voters in the voter set $n$.

\subsection{Partial Axioms}

The intuition for these axioms is unlike that of the voter groups in global axioms; in essence, it is fair for a voter group to deserve representation even if they have a non empty intersection of approval choices on at least some offices if not all. Therefore, partial axioms weaken the precedent of the condition for representation that is mandated by global axioms and allow even `almost cohesive' voter groups in that sense to have at least some kind of representation. For example, a group of voters that agree upon candidates in $3$ out of $5$ offices should not ideally be overlooked as a non-cohesive group and hence deserve some representation.  
\begin{definition}[Partial Justified Representation]
An office allocation $X$ satisfies \emph{Partial Justified Representation} if all voters in $V' \subseteq V$, have a non-empty intersection of approval ballots for at least $k'$ distinct offices and $\lfloor \frac{|V'|}{|V|} k' \rfloor > 0$ then, $((\cup_{v \in V'}v) \cap X) \neq \emptyset$.  
\end{definition}
Formally, if $i, l \in [k]$ and $\exists k', T = \{A_i, \dots A_l\}$ such that $(\cap_{v \in V'}v) \cap j \neq \emptyset, \forall j \in T$ and $|T| \geq k'$ and $\lfloor \frac{|V'|}{|V|} k' \rfloor > 0$ then $((\cup_{v \in V'}v) \cap X) \neq \emptyset$. 

\begin{example}{\label{example: partialJR}}
Suppose we have a scenario where $k = 5$ offices have to be filled by aggregating the preferences of $|V| = 100$ voters. In such a case, an office allocation $X$ satisfies \emph{Partial Justified Representation} if all groups of voters of size $|V'| = 50$ and have at least $2$ candidates in the intersection of their approval ballots get at least one of their preferred candidates in one of the offices. Similarly, each voter group of size $25$ agreeing on $4$ candidates should get satisfied and so on. 
\end{example}

\begin{observation}{\label{observation : observation1}}
Partial Justified Representation is a stronger proportionality notion than Global Justified Representation because note that the criterion for cohesiveness in Global Justified Representation is stricter than that in Partial Justified Representation. That is, GJR compels voter groups to agree on at least one candidate in every office; on the contrary, Partial JR does so for only $k' \leq k$ offices, such that the size of the voter group if at least $n/k'$. Therefore, any algorithm that produces an office allocation that satisfies Partial (Proportional) Justified Representation, satisfies Global (Proportional) Justified Representation. 
\end{observation}

\begin{proposition}{\label{proposition: PartialJR}}
An office allocation $X$ satisfying Partial Justified Representation always exists.
\end{proposition}
\begin{proof}
First of all, we traverse through the set $2^V$ and for each $V' \subseteq V$, we find the number of offices upon which they have a non empty intersection i.e. $k'$. Let the set of all groups of voters that deserve representation be $\mathcal{P}$. We define $\mathcal{P}$ as the set of all voter groups for whom $|V'| \geq n/k'$. Formally, 
$$\mathcal{P} = \{V' \subseteq V: |V'| \geq n/k'\}$$
Note that, all the groups of voters $V' \in \mathcal{P}$ have a size $\geq n/k$, since, 
$V' \geq n/k'$ and $k' \leq k$. Pick up voter groups $V'$ from the set $\mathcal{P}$ in 
order of increasing $k'$ (the number of offices on which the voter group has a non-empty
intersection) and allocate an office (out of the $k'$ offices) to one of the candidates 
belonging to the intersection of the approval ballots of the voters for that office. 
Eliminate all such voter groups $V''$ from $\mathcal{P}$ such that $V' \cap V'' \neq 
\emptyset$ and re-iterate. The algorithm terminates when $\mathcal{P} = \emptyset$. \\
For correctness, observe that after each iteration, the algorithm satisfies the voter set $V'$ along with all the other voter groups $V''$ such that $V' \cap V'' \neq \emptyset$ by allocating one office to one of the candidate who lies in the \emph{intersection} of the approval ballots of the voters of the voter group. Thus, due to the requirement of only one voter of every voter group to be satisfied the elimination of these voter groups works. This renders every voter group eliminated to be mutually disjoint; also, every voter set $V' \geq n/k$ and this allows the algorithm to allocate at most $k$ officers to the given offices. In effect, the algorithm is able to satisfy disjoint sets of voters, each of size $\geq n / k$. \\
Now we show that the algorithm does not reach a stage at which it runs out of appropriate offices to fill - i.e. those places on which the voters of the group have a non-empty intersection but are already filled. Say there exists a group of voters $V^*$ whose voters have a non-empty intersection on $k^*$ offices, but these are filled already by previously satisfied voters. Now, this means that at least $k^*$ groups of voters have been satisfied. Note that, the value of $k'$ in the algorithm is taken in increasing order and therefore, voter groups having non-empty intersection on some $k'' \leq k^*$ offices have already been satisfied. This means the sizes of these satisfied groups are at least the size of $V^*$ because $|V'|\cdot k' \geq n$ (as $k'$ increases, size of the voter group must decrease for conservation of the bound over the value of $n$). Due to the mutually disjoint nature of all these voter groups, the algorithm already took care of at least $k^*$ groups of size at least $|V^*|$. In turn, $|V^*| \geq n/k^*$ which implies that the number of voters already satisfied $\geq k^* \cdot \frac{n}{k^*} \geq n$. Thus, we arrive at a contradiction that there is some group of voters $V^*$ to be taken care of, when all the voters in question have been satisfied, hence proving that the existence of such a group of dis-satisfied voters is impossible.   
\end{proof}

Similarly, we define \emph{Partial Proportional Justified Representation} to strengthen the representation of groups of voters that are large enough but do not get representation proportional to their size.

\begin{definition}[Partial Proportional Justified Representation]
An office allocation $X$ satisfies \emph{Partial Proportional Justified Representation}(\textsc{Part-PJR}) if all voters of every voter group $V' \subseteq V$ have a non empty intersection of approval ballots on at least $k'$ distinct offices such that $\lfloor \frac{|V'|}{|V|} \cdot k' \rfloor > 0$ and $(\cup_{v \in V'} v) \cap X \geq \lfloor \frac{|V'|}{|V|} \cdot k' \rfloor$.
\end{definition}

In a situation analogous to Example \ref{example: partialJR}, in an office allocation satisfying \textsc{Part-PJR}, all voter groups of size $|V'| = 25$, who have $k' = 8$ candidates in the intersection of their approval sets should receive at least $2$ of their preferred candidates to hold some offices and so on. \\
We now prove that an office allocation satisfying Partial Proportional Justified Representation exists using a constructive proof. The algorithm that we propose for the corresponding proof is fixed-parameter tractable wrt. the number of voters $n$ but runs in super-polynomial time. 

\begin{proposition}
An office allocation satisfying Partial Proportional Representation always exists. 
\end{proposition}
\begin{proof}
The existence proofs follows in two steps namely, deciding whom to satisfy and how to satisfy respectively. To this end, in the first step we find $\mathcal{P} = \{V' \in v: |V'|k' \geq |V|\}$ where $k'$ is the number of offices on which the voters have a non-empty intersection. In each iteration, we select a $V' \in \mathcal{P}$ which has the highest value of $|V'|k'$ since these groups are the most `inclusive' of similarly thinking voters. While iterating over each $V' \in \mathcal{P}$, we maintain a set of conditions $\mathcal{C}$ (which is initially empty) and add a condition requiring the filling of $\lceil \frac{|V'|}{|V|}k' \rceil$ offices of the $k'$ offices on which voters of $V'$ have a non-empty intersection. Further, we prune the set $\mathcal{P}$ by removing the group $V'$ and all other $V''$ such that $V' \cap V'' \neq \emptyset$ so as to ensure already satisfied voters are eliminated from further consideration and reiterate. The first phase comes to an end whenever $\mathcal{P}$ is empty. \\
In the second phase of the algorithm, we enforce the conditions in the order of increasing $k'$ (observe that every condition $c \in \mathcal{C}$ has a corresponding $V'$, which in turn has a corresponding $k'$, satisfying the condition $k' \geq |V|/|V'|$. \\
For correctness, we show that there always exists an office that can be filled. Furthermore, it is important to show that there always exist enough offices to satisfy voters of all $V'$'s such that the corresponding conditions ($c \in \mathcal{C'}$) of filling offices can be satisfied without all these offices being filled already to satisfy other voters considered in the previous iterations. Assume towards a contradiction that during the run of the algorithm, we encounter a voter group $V^*$ that have an intersection on $k^*$ offices but the offices have already been filled. Since all these offices have been filled, it means that the algorithm explicitly has taken care of at least $k^*$ voter groups, each being a disjoint set of voters having intersection in at least $k' \geq k^*$ offices (since we consider office allocation in increasing order of $k'$). Thus, $k'$ voter groups of size $|V'| \geq V^*$ have already been satisfied (since $k'$ increases and $|V'|*k' \geq |V|$, $|V'| \geq V^*$). Further, it is important to note that $|V^*| \geq |V|/k^*$, essentially because it has been satisfied and it must have been deserving. This means that the number of voters that have already been taken care of are $k^* \cdot |V|/k^*$ = $|V|$, which contradicts the fact that a set of voters $V^*$ exists that contains unsatisfied voters. Therefore, there doesn't exist such a set and all voters are satisfied by the algorithm.
\end{proof}

\section{On Unequal Importance}

It is important to note that the basic assumption till now has been that every office in the proposed executive wields an equal amount of power. However, this assumption is seldom true in the context of the governments all around the world. Therefore, considering the extent of power that every office exercises is an important fairness influencing parameter while designing algorithms for office allocation. For example, if there is an election for deciding office allocation for the offices of Defense, Health, Transport and Housing development, it should be noted that even if a cohesive voter group forms 75\% of the electorate, it is not justified to elect 3 out of the 4 candidates to any $3$ offices that the voter group demands because of the inherently different amounts of power held by every office.  \\
In such cases, measuring the amount of power that any office wields becomes the base for design of fair algorithms for electing the executive.

\begin{definition}[Objective Power Function]
A power function $\mathbf{P^{obj}} : O \times V  \mapsto \mathbb{R}$ is said to be an \emph{Objective Power Function} if every voter believes the concentration of power at every particular office to be equal. That is, $\mathbf{P_i^{obj}}(o) = t_o \geq 0$, such that, $t_o$ is a constant, $\forall i \in V$ and $o \in O$.
\end{definition}

\begin{example}{\label{example : objective}}
Let's say that a government has the previously mentioned 4 offices to be filled namely, Defense, Health, Transport and Housing Development and that there are 4 voters voting for the election of candidates to these seats. One possible objective power function, $\mathbf{P^{obj}}$ for such a scenario can be specified as follows:
\begin{center}
    \begin{tabular}{ c c c c c }
         &  Defense & Health & Transport & Housing Dev. \\
        $v_1$ & 8 & 5 & 2 & 1 \\
        $v_2$ & 8 & 5 & 2 & 1 \\
        $v_3$ & 8 & 5 & 2 & 1 \\
        $v_4$ & 8 & 5 & 2 & 1 \\
    \end{tabular}
\end{center}
\end{example}
Clearly, every voter has the exact same perception of power that every office wields, i.e. the objective power function exercises dependence only on the office $o \in O$ and is independent of voters. Therefore, if the voter $v_1$ gets one of her favourite candidate in the defense office, she would be controlling an influence equivalent to 8 units. However, a more general power function, i.e. the Subjective Power Function can be defined as follows. 

\begin{definition}[Subjective Power Function]
A power function ~$\mathbf{P}^{sub} : O \times V \mapsto \mathbb{R}$ is said to be a \emph{Subjective Power Function} if every voter believes that the concentration of power is different with different offices. Formally, $\mathbf{P_i}^{sub}(o) \geq 0$, $\forall i \in V, \forall o \in O$.
\end{definition}

\begin{example}{\label{example: subjective}}
Say, there are two voters $\{v_1, v_2\}$ and two offices to be filled, $\{o_1, o_2\}$, the candidates for each of these being $\{a, b, c\}$ and $\{d, e, f\}$ respectively. Suppose that for $v_1$, office $o_1$ has a power $5$ and $o_2$ has a power $10$ and that for voter $v_2$, its the exact opposite. If $v_1$ and $v_2$ have approval sets $\{a, d, e\}$ and $\{b, e, f\}$ respectively, then we can translate these choices to cardinal utilities and say that $u_{v_1}(a) = 5$ (the power of office $o_1$) and $u_{v_1}(d) = 0$, $u_{v_1}(e) = u_{v_1}(f) = 10$. Practically, a Realtor would definitely feel that the influence of the policies that the Housing Development office has on her outweighs the influence that the Transport office makes. On the other hand, for a practicing doctor, the policies of a health office could be much more important than the defense office policies. 
\end{example}

\begin{remark}
It is noteworthy that Objective Power setting that we present are a case of restricted domain of our Subjective Power setting. Formally, $\mathbf{P_i^{sub}}(o) = t_o$, $\forall i \in V$, such that $t_o$ is a constant $\forall o \in O$. 
\end{remark}

\begin{remark}
The use of approval ballots in a sense acts as a way for voters to elicit their preferences without much cognitive burden. In such settings, expecting voters to estimate subjective power functions would be tedious and therefore unnatural. However, a natural assumption to make in this direction would be that there exists an approximately equal perception amongst voters about the relative powers wielded by different offices which makes objective power functions model voter preferences more accurately. Therefore, it makes sense to study proportionality in office allocations where voters objectively define powers for the offices in question. 
\end{remark}

\subsection{Proportional Objective Power Allocation}

A natural way (in adherence to Justified Representation) to realize proportional allocation of objective power to voters would be to grant an office to a sufficiently large group of cohesive voters which unanimously approve it. Without loss of generality, assuming that no voter is allowed to submit an empty ballot for any of the departments, the following axiom formalizes a strong notion for proportional allocation of power:

\begin{definition}[Most Important Power Allocation]
An office allocation $X$ is said to have a \emph{Most Important Power Allocation} if $\forall V' \subseteq V$, there exists an office $o \in O$ such that $eligible(o) \subseteq \cap_{i \in V'} A_i$ and $|V'| \geq \mathbf{P^{obj}}(o) \cdot n / \Sigma_j\mathbf{P^{obj}}(j)$ then, $(X \cap eligible(o) \cap (\cup_{i \in V'}A_i)) \neq \emptyset$.     
\end{definition}
In other words, if a group of 
However, there does not always exist an office allocation that possesses a most important power allocation. This can be shown using the following example.

\begin{example}\label{example : failMIPA}
Consider an electorate consisting of $10$ voters, $2$ offices $o_1$ and $o_2$ of powers $5$ each. Assume that any $5$ voters of the electorate approve a candidate $a$ on the office $o_1$ while the remaining $5$ voters approve a candidate $b$ on the same office and that there is no such consensus on the office $o_2$. In such a case, at least one of the two suitably large and similarly inclined group of voters looses out on expression of choice since only one candidate of $a$ and $b$ can occupy office $o_1$. 
\end{example}

\subsection{Weak Most Important Power Allocation}
A weaker version of Most Important Power Allocation would be to mandate at least one unit of power to every deserving and cohesive voter group, even though it disproportionate to the group's size. The following definition for \emph{Weak Most Important Power Allocation} formalizes such a notion.

\begin{definition}[Weak Most Important Power Allocation]
An office allocation $X$ is said to have a \emph{Weak Most Important Power Allocation} if for all groups of voters $V' \subseteq V$, if there exists an office $o \in O$ such that $eligible(o) \subseteq \cap_{i \in V'} A_i$ and $|V'| \geq \mathbf{P^{obj}}(o) \cdot n / \Sigma_j\mathbf{P^{obj}}(j)$ then, $((\cup_{i \in V'} A_i) \cap X) \neq \emptyset$.
\end{definition}

We can satisfy this weak axiom by a simple polynomial time algorithm GreedyCC.

\subsubsection{GreedyCC}
The algorithm advances in iterations as follows: in each iteration, find the candidate which has the highest approval score which is at least equal to $\mathbf{P^{obj}}(o)\cdot n / \Sigma_o \mathbf{P^{obj}}(o)$, where $o$ is the office to which the candidate belongs. Designate that candidate to its corresponding office and remove the satisfied voters from further consideration and continue to the next iteration. If at any point in the algorithm, there are no candidates left for which the approval score is strictly less than $\mathbf{P^{obj}(o)}\cdot n / \Sigma_o \mathbf{P^{obj}}(o)$ (where $o$ is the office to which the candidate belongs), randomly assign candidates to these remaining offices and terminate. \\
For correctness, observe that at every iteration of the algorithm, for an office which is filled, $\geq \mathbf{P^{obj}(o)}\cdot n / \Sigma_o \mathbf{P^{obj}}(o)$ distinct voters are satisfied. Therefore, after filling all the possible offices $o \in O$, the total number of voters that shall be satisfied will be $\geq \Sigma_{i \in O}\mathbf{P^{obj}(i)}\cdot n / \Sigma_o \mathbf{P^{obj}}(o) = n$.


\section{Hierarchy of Axioms and Preference Settings}
\begin{figure}[htp]
    \centering
    \includegraphics[width=10cm, height=6cm]{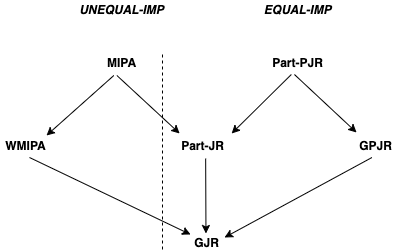}
    \caption{Hierarchy of axioms}
    \label{fig:axioms}
\end{figure}

\begin{figure}[htp]
    \centering
    \includegraphics[width=10cm, height=6cm]{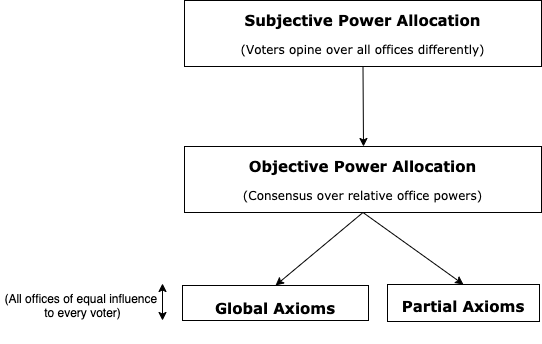}
    \caption{Hierarchy of Preference Settings}
    \label{fig:settings}
\end{figure}

In this section, we present the hierarchy of axioms and preference elicitation settings proposed in the paper. Figure \ref{fig:axioms} shows the hierarchy of the strength of axioms and Figure \ref{fig:settings} reflects a hierarchy of voter expressivity / domain restriction scheme that we study.   

\section{GreedyPAV}

In this section, we first describe the voting rule GreedyPAV and establish its utility using simple toy examples.  We examine the capability of this voting rule in satisfying our axioms when the preference profiles of the voters are synthetically generated using two commonly used models: the Impartial Culture Model and the Polya-Eggenberger Urn Model \cite{BERG1985271, polya, walsh2011hard}. Further, we discuss a natural setting in the election of executive which calls for the design of an iterative construct for preference aggregation. Finally, we provide a socially explainable Paper-and-Pencil Realization of greedyPAV which works efficiently for our original settings and the one motivated in sub-section \ref{section:declined}.

\subsection{Adapting GreedyPAV}

GreedyPAV is used for multiwinner elections and is known to be proportional for that setting. In those settings, it works as follows:
  Initially, each voter has a weight of $1$; the rule works in $k$ iterations (as the task in standard multiwinner elections is to select a set of $k$ alternatives), where in each iteration one alternative will be added to the initially-empty committee.
  In particular, in each iteration, the alternative with the highest total weight from voters approving it is selected, and then the weight of all voters who approve this alternative is reduced; the reduction follows the harmonic series, so that a voter whose weight is reduced $i$ times will have a weight of $1 / (i + 1)$ (e.g., initially the weight is $1$; then, a voter reduced once would have a weight of $1/2$, then of $1/3$, and so on).

In the proposed adaptation of GreedyPAV to our setting of electing an executive branch, in each iteration, we again select the alternative with the highest weight from approving voters; say this is some $a_j \in A_j$. Now, we fix the $j$th office to be populated by $a_j$; then, as it is fixed, we remove all other $a_i \in A_j$ from further consideration (as the $j$th office is already populated) and reweight approving voters as described above (in the description of GreedyPAV for the standard setting of multiwinner elections).

\begin{example}
Consider again the election described in Example~\ref{example:toy}, consisting of $k$ sets of alternatives: $A_1 = \{a, b\}$, $A_2 = \{c, d\}$, and $A_3 = \{e, f, g\}$; $3$ voters: $v_1 = \{a, c, e\}$, $v_2 = \{a, c, f\}$, and $v_3 = \{a, d, f\}$.

In the first iteration of GreedyPAV, we will select alternative $a$ to populate the first office; then we reweight all votes to be $1/2$ (as all voters approve $a$). In the next iteration we will select either $c$ or $f$ (as both has total weight of $1$); say that our tie-breaking selects $c$.\footnote{we omit discussion on tie breaking as it technically clutters the presentation; say that we do it arbitarily following some predefined order over all alternatives.} Then, we reweight $v_1$ and $v_2$ to be both $1/3$. In the third and last iteration, alternative $e$ has $1/3$ weight, while alternative $f$ has $1/3 + 1/2$ so we select $f$. Thus, GreedyPAV assigns $a$ to the first office, $c$ to the second office, and $f$ to the third office.
\end{example}
\subsection{Experimental Analysis: GreedyPAV}

We study the efficacy of the GreedyPAV rule to elect an office allocation that satisfies our axiom Global Justified Representation; this is so because GJR as it were can be considered to be the closest naturalisation of the Justified Representation axiom in Multi-Winner Elections to our settings. 

\subsubsection{Experimental Setup}
In particular, we generate approval ballots for voters uniformly at random (the Impartial Culture model) and by varying the value of $\alpha$, which imitates thought cohesivity amongst voters in  the Polya-Eggenberger urn model. Higher is the value of $\alpha$, more is the thought cohesivity amongst voters and the more similar their ballots look. After generating these preferences, we input these to the GreedyPAV procedure and check if the output of the rule satisfies GJR. In one epoch, we generate 100 and 200 varied voter profiles (and 500 instances too for Impartial Culture Model) and find the probability with which the output of the function satisfies GJR and tabulate our results in tables \ref{tab:IC} and \ref{tab:UrnGJR}.

\subsubsection{Discussion and Analysis}
As observed from the tables \ref{tab:IC} and \ref{tab:UrnGJR}, the probability that the GreedyPAV rule elects a committee that satisfies a basic proportionality criterion, GJR is 75 \% to 85\%. The outliers in this range are those values that are a result of greater similarity amongst approval ballots of voters due to higher values of $\alpha$ in the Urn Model Simulations. The increase in the number of instances does not seem to have a particularly uniform effect on the efficiency of GreddyPAV to elect proportional office allocations, though urn model simulations point toward an increasing efficiency trend with an increase in the number of instances.
The efficiency of GreedyPAV to elect proportionally representative office allocations is reasonable given the additional benefits it fields in comparison to other voting rules; for e.g. it is much easier to handle candidates declining seats and is easily explainable (as discussed in sections \ref{section:declined} and \ref{section:paper}).   

\begin{table}
    \centering
    \begin{tabular}{|c|c|}
    \hline
    Instances     & GJR  \\
    \hline
    100     & 0.79 \\
    200 & 0.805 \\
    500 & 0.752 \\
    \hline
    \end{tabular}
    \caption{Probabilities for IC Model of Preference Generation}
    \label{tab:IC}
\end{table}


\begin{table}
    \centering
    \begin{tabular}{|c|c|c|c|c|c|}
    \hline
    Instances & $\alpha=0.1$ & $\alpha=0.3$ & $\alpha=0.5$ & $\alpha=0.7$ & $\alpha=0.9$ \\
    \hline
    100 & 0.7 & 0.78 & 0.8 & 0.8 & 0.84\\
    200 & 0.743 & 0.85  & 0.88 & 0.87 & 0.885 \\
    \hline
    \end{tabular}
    \caption{GJR satisfying Probabilities for Polya-Eggenberger Urn Model of Preference Generation}
    \label{tab:UrnGJR}
\end{table}

\subsection{Declined Candidates}\label{section:declined}

We briefly discuss how to deal with candidates who are selected to an office but decline to serve in the office:
  In particular, assume that, for a given instance, there is a candidate $c$ that is selected to some office $A_j$ as a winner; however, when the day comes, $c$ refuses to populate the $j$th office (say, e.g., that $c$ accepts a different career).

A simple solution would be to simply run the aggregation method again, after removing $c$ from the election. However, when using greedyPAV, it might be the case that, as a result, other offices will get different candidates as winners. As this might be unacceptable (as it means, e.g., that if the foreign minister declines to accept then we change the environmental minister), we offer a different option, as follows.

In particular, we can keep the other $k - 1$ ministers intact, remove $c$ from the election, and run a single further iteration of greedyPAV, resulting in a different candidate to be selected for the office that $c$ was originally elected for. This ensures that the other winners are kept as they were, while the weights of all voters are calculated properly, and a different candidate is being selected for that office.

An unexpected beneficial byproduct of this method is that the replacement of an impeached minister is not known in advance, before calculating it from the ballot. This reduces the motivation of any particular candidate or their supporters to initiate an unjustified impeachment or recall elections for a minister.

\subsection{Paper-and-Pencil Realization}\label{section:paper}
An essential property of a voting rule is that it can be easily explained to  the voter.  Another important property is that its realization does not require trusting external elements (e.g., hardware and software).  Fortunately, greedyPAV was invented before computers and hence must have been realized initially without them.

For completeness, clarity, and ease of implementation, we describe here a pencil-and-paper
realization of our greedyPAV protocol for electing the executive branch, including for determining replacements for declined candidates.

The basic process is as follows:
\begin{enumerate}
    \item Before the vote commences, there is a finite list of candidates and a finite list of voters. Each candidate name is associated with one office.
    \item During the vote, every voter writes a list of names on a note, places the note in an envelope and then in the ballot box.
    \item All envelopes are collected and opened. If there is a limit on the number of names a voter can vote for, then all excess names on a note, as well as names of non-candidates, are stricken with X's.  If any name in the note is stricken with a line, then it is stricken again with X's.
    \item The \emph{weight} of a name in a note is $1/(k+1)$, where $k$ is the number of names stricken with a line in the note, if the name appears in the note, and zero otherwise.
    \item Before vote counting commences, all offices are vacant and no name on any note is stricken with a line.
    \item Counting proceeds in rounds until all offices are occupied (or no vacant office has a candidate named in a note) as follows: In each round, the combined weight of each name in all notes is computed.  The highest-weighted name for a vacant office is elected, occupies its office, and the name is stricken with a line from each note it appears in.
\end{enumerate}
This completes the description of the basic voting process.  In case an elected minister cannot fill her office, a replacement is needed (as described    in Section~\ref{section:declined}).  
To realize the concept described there (in Section \ref{section:declined}), the following simple procedure is followed:
The name of the declining minister-elect is stricken from all notes with X's,  the highest-weight candidate for the vacated office is elected, and her name is stricken with a line from all notes.  The resulting notes can be kept for calculating any future replacements.

\section{Outlook}

We have described the setting of selecting the executive branch via direct democracy. For this setting we suggest the use of an adaptation of GreedyPAV and show, via computer-based simulations, that it indeed does not disregard minorities in many cases.\\
Domain restriction over different offices can be used to inspect if some of the algorithms become polynomial time for electing a proportionally representative executive. \\
There is an important temporal equivalent of the `unequal importance' case; for example, electing Presidents for a country across years when the importance of different years are different for voters. \\
It would also be exciting to study what happens when candidates decline seats., a toy example of a fundamental impossibility has been mentioned in section \ref{subsubsection: AcrossMinistries}.

\subsection{Participation Across offices}{\label{subsubsection: AcrossMinistries}}
The current settings for the election of executive do not allow a candidate to contest for more than one office. A natural extension of our settings therefore would be to allow voters to nominate a candidate to more than one offices. The application of greedyPAV (GreedyPAV) to such settings however, fails to take into account the choices of even half of the populace.

\begin{example}
Let's say, there are four voters $v_1, v_2, v_3, v_4$, 2 offices $o_1, o_2$ and two candidates say, $a, b$. Denote the choices of the voters for $o_1$ and $o_2$ respectively by: 
\begin{align*}
    v_1, v_2 : \{\{a\}, \{b\}\} \\
    v_3, v_4 : \{\{b\}, \{a\}\}
\end{align*}
Assuming lexicographic tie-breaking, greedyPAV would elect $a$ to the office $o_1$ and $b$ to office $o_2$. Therefore, voters $v_3$ and $v_4$, which form exactly 50\% of the total electorate would be left dissatisfied altogether. 
\end{example}
The above example can be easily realised in practical situations; for instance, if a voter nominates some candidate for the Department of Education but she gets elected to the Department of Defense, she would not be satisfied.

\bibliographystyle{plain}
\bibliography{bib}

\end{document}